\documentclass[mathleft, fleqn, preprint]{an}
\usepackage{graphicx,epstopdf}
\usepackage{times}
\usepackage{amsmath}
\usepackage[utf8]{inputenc}
\usepackage[russian, english]{babel}
\usepackage{soul}
\begin{document}

\Pagespan{789}{}
\Yearpublication{2016}%
\Yearsubmission{2015}%
\Month{11}%
\Volume{999}%
\Issue{88}%

\title{Long-term variations in sunspot magnetic field -- area relation}

\author{Yury A. Nagovitsyn\inst{1},
Alexei A. Pevtsov\inst{2}\fnmsep\thanks{Corresponding author:
  \email{apevtsov@nso.edu}\newline},
\and  Aleksandra A. Osipova\inst{1}
}
\titlerunning{Long-term variations in sunspot  magnetic field -- area relation}
\authorrunning{Yu.A. Nagovitsyn, A.A. Pevtsov \& A.A. Osipova}
\institute{
Pulkovo Astronomical Observatory, Russian Academy of
Sciences, Pulkovskoe sh. 65, St. Petersburg, 196140 Russian Federation
\and 
National Solar Observatory, Sunspot, NM 88349, U.S.A.
}

\received{}
\accepted{}
\publonline{later}


\abstract{%
Using observations of sunspot magnetic field strengths ($H$)  from the Crimean Astrophysical Observatory 
(CrAO) and area ($S$) of sunspots from the Kislovodsk Mountain Astronomical Station of Pulkovo 
Observatory, we investigate the changes in the relation between $H$ and $S$ over the period of about two
solar cycles (1994--2013). The data were fitted 
by  $H = A + B \log S$, where $A = (778\pm46)$ and $B = (778\pm25)$. We show 
that the correlation between $H$ and $S$ varies with the phase of solar cycle, and
$A$ coefficient decreases significantly after year 2001, while $B$ coefficient 
does not change significantly. Furthermore, our data confirm the presence of two distinct 
populations in distribution of sunspots (small sunspots with weaker field strength and large sunspots 
with stronger field). We show that relative contribution of each component 
to the distribution of sunspots by their area 
changes with the phase
of solar cycle and on longer-then-cycle periods. We interpret these changes as 
a signature of a long-term (centennial) variations in properties of sunspots.
}

\maketitle

\section{Introduction}

Cycle 24 exhibited several anomalous properties such as unusually low and prolong minimum between cycles 23 and 24, strong hemispheric 
asymmetry in sunspot activity, low amplitude as compared with more recent cycles, not-well organized polar field reversal in the Northern hemisphere (see Otkidychev and Skorbezh, 2014 and references therein). These properties 
stimulated number of recent 
studies on possible long-term variations of physical properties of 
sunspots (Pevtsov et al. 2011, 2014; Nagovitsyn et al. 2012; Livingston, Penn \& Svalgaard 2012; 
Rezaei et al. 2012; DeToma et al. 2013). Based on routine observations of sunspots in 
spectral line Fe {\rm I} 1564.8 nm., Penn and Livingston (2007, 2011)
concluded that average field strength of sunspot magnetic field monotonically declined over 1998--2011. 
Newer observations seem to confirm that trend (Watson et al. 2014). The trend was
interpreted as an indication of a global decline in sunspot activity that may herald the Sun entering 
a grand Maunder-type minimum (Livingston et al. 2012). 

In contrast, using synoptic 
observations of sunspot field strengths taken from 1957--2011 in the framework of ``Sun Service'' program 
(\foreignlanguage{russian}{Служба Солнца}, in Russian), 
Pevtsov et al. (2011) concluded that  yearly-average field strengths of 
strongest sunspots show cycle variations, and that the 1998--2011 declining trend reported by 
Penn and Livingston (2006, 2011) coincides with the declining phase of cycle 23. 
Pevtsov et al. (2014) extended early findings of Pevtsov et al. (2011) to the 1920--1958 period (using observations from
Mount Wilson Observatory). The data showed clear solar cycle variations, and there was no notable long-term trend.

In addition to direct measurements of magnetic fields, one can estimate sunspot field strength from their areas using a known relation between the area of sunspots and the field strength (e.g., Ringnes \& Jensen 1960; Pevtsov et al. 2014; Tlatov \& Pevtsov 2014; Mu{\~n}oz-Jaramillo et al. 2015).
Nagovitsyn et al. (2012) showed
that both the gradual decline in average field strengths (observed by Penn \& Livingston 2006) and the 
cycle variation in field strength of strongest sunspots (observed by Pevtsov et al. 2011) can be explained
by changes in the distribution of sunspots. They showed that the distribution of sunspots by their area is bimodal, with one component 
corresponding to ``small" and the other to ``large" sunspots with peak distributions at about 17 MSH 
(millionth of solar hemisphere) and 174 MSH, accordingly. They interpreted such  bimodal distribution as an 
indication that sunspots of different size may originate from different depths in the solar convection zone. 
The overall decline in averaged field strength can be 
explained by change in relative contribution of these two distributions ({\it i.e.,} larger fraction of small spots 
as compared with large  spots will result in a smaller average field strength). 
DeToma et al. (2013) reported a deficit of large sunspots in cycle 23 as compared with some of the previous cycles.
The presence of two components in the distribution of sunspot areas was recently confirmed by Mu{\~n}oz-Jaramillo et al. (2015),
who found that the distribution of sunspots by area can be fitted by a combination of the 
Weibull distribution (representing contribution of
 small sunspots), and the log-normal component (representing contribution of large spots). 
Cho et al. (2015) confirmed the earlier findings that pores, transitional and mature sunspots form different functional 
dependencies in magnetic field -- area relation. They also found that the distributions of umbral areas of transitional and mature 
sunspots exhibit distinctly different properties.

Tlatov \& Pevtsov (2014) reported near linear relation between the logarithm of area of sunspots $S$ and
their maximum magnetic field strength $H$ based
on observations from the Helioseismic and Magnetic Imager (HMI) on the Solar Dynamics Observatory (SDO). 
Mu{\~n}oz-Jaramillo et al. (2015) found that sunspot magnetic flux $\Phi$ and areas $S$ are best fitted by a power--law function
$\Phi = (1.95\pm0.14)\times 10^{19} \times S^{(0.98\pm0.01)}$, where $\Phi$ is in units of Mx and $S$ is in MSH. Fitting the 
$\Phi$ and
umbral areas returned slightly different fits for SDO/HMI data
$\Phi = (5.20\pm0.03)\times 10^{19} \times S^{(1.08\pm0.01)}$ and SOHO/MDI data
$\Phi = (6.21\pm0.11)\times 10^{19} \times S^{(0.97\pm0.01)}$. Unlike space based data, ground based observations
are affected by atmospheric seeing. Furthermore, many past (historical) measurements provide areas of whole sunspots
(not umbral areas). This may result in a different functional dependency between sunspot field strengths and their areas.
 Pevtsov et al. (2014) used $H$ measurements from MWO and $S$ observations 
from Royal Greenwich Observatory (RGO) to fit $H = (-774.2\pm35.6) = (536.0\pm7.7)\times\ln(S)$.

This article aims at verifying and further investigating various tendencies related to sunspot magnetic 
fields and their areas.

\section{Data}
\label{sec:data}

For the sunspot areas we employed data from the Kislovodsk Mountain Astronomical Station (KMAS, www.solarstation.ru) 
of the Central Astronomical Observatory at Pulkovo. For sunspot field strengths we used measurements from Crimean 
Astrophysical Observatory 
(CrAO, solar.craocrimea.ru/eng/observations.htm) taken from 1994--2013. Sunspot field strengths were measured 
manually  via the separation between the two Zeeman components of a magnetic-field-sensitive spectral line as described
in Pevtsov et al. (2011). The observer identified the location in sunspot umbra with the largest separation between the Zeeman components and measures the separation between them. Thus, in the following discussion, we referred to these measurements as maximum field strength. Additional details about sunspot field strength measurements at Crimean Astrophysical Observatory can be found elsewhere (Lozitska et al. 2015). Sunspot areas were determined using daily photoheliograms taken 
with a broadband filter in visible range. The areas are then corrected for foreshortening by dividing an area in the image plane by the cosine of heliocentric angle of the sunspot 
center. This simple transformation may underestimate the curvature of the solar surface for features situated close to solar limb. However, the errors are minor given the relative size of typical sunspots relative to the size of solar sphere. 

By their cumulative distribution, KMAS group areas are similar to RGO data (Mu{\~n}oz-Jaramillo et al. 2014)
although there appears to be a minor depletion of the smaller groups in KMAS data. Still, based on additional tests 
of RGO and KMAS data taken for the same period of time (1955--1976) we find that 
the bimodal distribution is present in both RGO and KMAS data, and thus,
this minor depletion does not 
affect the presence of bimodal distribution in areas of sunspot groups.

Unlike RGO data, KMAS observations included information on areas of the main sunspot in each group.
The latter allows a direct comparison between the area and field strength in the main spot of each group.
The selection of the main sunspot of each group further emphasizes the contribution of larger sunspots. 
This is unavoidable, as the KMAS dataset does not contain area measurements for other sunspots in the group. 
Despite this limitation, the resulting subset is uniform in its statistical properties, and thus, is well-suited for a study 
of area-field strength relation and its change with the solar cycle.

Using  KMAS (sunspot areas) and CrAO (sunspot magnetic fields) data, we created two separate subsets. 
The first subset, $HS1$, contains 1767 pairs of $H$ and $S$ (total area including 
both umbra and penumbra) for the main sunspot in each group 
observed during 2012--2013. The second data set, $HS2$, contains 653 sunspots observed during 1994--2013 near solar disk center (heliocentric angle $\theta \le$ 14$^\circ$, or $\cos\theta>0.97$). Selection of sunspots near disk center minimizes the contribution of horizontal fields to the measured field strengths.
Disk passage of selected well-developed sunspots shows that the field strength changes as the function
of heliocentric distance. This change is not as steep as a cosine function, and within the $\approx \pm$
30 degree interval there is very little change. Thus, our $HS2$ data set does not require correction
of magnetic field strength for center-to-limb variation.

The measurements used in this study are the subject of scattered light. In general, the negative effects of the scattered light  
are more significant for sunspots situated closer to solar limb (as compared to near-disk center areas), and they are larger 
for small sunspots. The data are also the subject of the observer's bias, and the variations in the atmospheric seeing effects (e.g., due to scintillations). It is impossible to correct the synoptic historical data such as used in this article for the atmospheric seeing conditions. However, comparative analysis of statistical properties of sunspot field strength measurements suggests that the
magnetic field measurements  above 1100 Gauss in CrAO data are close to ``true'' field strengths (Lozitska et al. 2015).

For analysis of long-term trends we also employed combined RGO--SOON (Solar Observing Optical Network) dataset as described in Pevtsov et al. (2014).
Unlike KMAS observations, RGO data provide the total area of active regions only. Still, the total area of an active region
correlates well with the maximum field strength in main sunspot (Mu{\~n}oz-Jaramillo et al. 2015), and thus, use of RGO data 
in studies of $H$ vs. $S$ is justifiable. To combine RGO and SOON data into a single data set, we used the scaling coefficient determined by previous investigators (Hathaway 2010).

\section{Functional Dependence of Field Strength vs. Sunspot Area}
\label{sec:func}

Early studies (Nicholson 1931; Houtgast \& Sluiters 1948; Ringnes 1965) discussed various
functional dependencies between sunspot area and the maximum field strength including:

\begin{equation}
 H = A + B \times \log S
\label{eq:H-log}
\end{equation}
\begin{equation}
\log H = A_1 + B_1 \times \log S
\label{eq:log-log}
\end{equation}
\begin{equation}
H={{A_2 \times S} \over {B_2+S}}
\label{eq:H-S}
\end{equation}

Mu{\~n}oz-Jaramillo et al. (2015) fitted $H = a \times S^b$, which is similar to Equation \ref{eq:log-log}.
Equation \ref{eq:H-S} is a modified version of one used by Houtgast \& Sluiters (1948) with a new term added, which 
represents minimum field strength in sunspot.
Ringnes (1965) fitted Equation \ref{eq:log-log} to MWO observations from 1917--1956 and found that A$_1$ and B$_1$ 
coefficients exhibit long-term variations. Pevtsov et al. (2014) showed that value
of B$_1$ coefficient correlates with 
amplitude of solar cycle. Since this coefficient represents steepness of $\log H$ vs. $\log S$ dependence, Pevtsov et al. (2014) suggested that cycle dependence of this coefficient can be explained by changes in fraction 
of small and large sunspots in each cycle. Tlatov \& Pevtsov (2014) applied Student F-test to evaluate goodness of fit 
of magnetic field strength vs. sunspot areas (observed by SDO/HMI) by Equations \ref{eq:H-log} and \ref{eq:log-log}. Both functional dependencies were found to represent the data equally well. 
On the other hand, Ringnes \& Jensen (1960) concluded that Equation \ref{eq:H-log} provides the best 
representation of $H$ vs. $S$ dependence.
Currently, there is no physical model that can justify the use of a specific $H$ vs. $S$ functional dependence although Tlatov \&  Pevtsov (2014) noted that Equation \ref{eq:log-log} can be derived from the distribution of 
magnetic field of a dipole situated at a certain depth below the photosphere.

Following Pevtsov et al. (2014) for the following analysis we chose
to use Equation \ref{eq:H-log} to represent $H$ vs. $S$ dependence.
Table \ref{tab:t-correlation} shows Pearson correlation coefficients (r$_{\rm P}$) between $H$ and $\log S$. 
In that test, we sort
the data in the increasing order of heliocentric distances, $\theta$ and divide the data set on groups of 150 
points (pairs of areas and
field strengths). For each group we computed the mean heliocentric angle $\theta$ and r$_{\rm P}$.
All correlations are 
statistically significant at the 0.04  level for bins close to the disk center, and at the 0.07 level for bins near the limb.
For $\theta \le $ 59$^\circ$ mean
correlation
coefficient is about 0.8 (Table \ref{tab:t-correlation}). For $\theta >$ 59$^\circ$ r$_{\rm P}$
declines rapidly to about 0.43. Such rapid decline in correlation coefficient may be the results of several 
factors that depend on heliocentric distance: change in contribution of
horizontal and vertical components of magnetic field in sunspots, decrease in apparent depth of
sunspot (effect of Wilson depression), increase in errors in measurements of magnetic fields, and increased effect of
scattered light on magnetic field measurements (see further discussion in Section 4).
As an additional test, we used
observations (solar disk passage) of several large sunspots from the {\it Vector Stokes Magnetograph} (VSM)
on {\it Synoptic Long-term Investigations of the Sun} (SOLIS) facility
(Balasubramaniam \& Pevtsov 2011). The magnetic field strength in these sunspots was determined
using Zeemanfit code (Hughes et al. 2013), which fits a combination of Gaussian and Voigt
functions to the observed Stokes I profiles. The Zeemanfit approach was developed to imitate
the manual measurements of magnetic field strengths similar to ones employed at CrAO. The results of 
this test show a decline in measured field for sunspots situated closer to solar limb as compared with their 
position near disk center. The latter agrees with the observed changes in coefficients $A$ and $B$ fitted to
Equations \ref{eq:H-log} (Table \ref{tab:t-correlation}). Coefficient $B$ presents the steepness of
$H$ vs. $\log S$ relation.
Increase in $A$ together with decrease in $B$ suggests that for the same size of
sunspots the magnetic fields are weaker for sunspots located at large $\theta$ (closer to the limb)
as compared with sunspots situated near disk center.

\begin{table}
\caption{Correlation coefficients, r$_{\rm P}$ between $H$ and $\log S$ for different heliocentric angles $\theta$, and $A$, $B$ coefficients fitted to Equation \ref{eq:H-log}. 
 To emphasize 
the
change in correlation (and in fitted $A$ and $B$ coefficients for functional
representation by Equation \ref{eq:H-log}), entries for larger $\theta$ are shown in bold font.}
 \begin{tabular}{rcrc}
\hline
$\theta$ Range&Correlation Coefficients&\multicolumn{2}{c}{Coefficients in  Eq. \ref{eq:H-log}}\\
degree&r$_{\rm P}$&\multicolumn{1}{c}{A}&B\\
\hline
3–-17&0.85&576$\pm$76&788$\pm$41\\
17–-22&0.79&631$\pm$94&776$\pm$49\\
22–-26&0.80&701$\pm$82&727$\pm$45\\
26–-31&0.80&646$\pm$91&773$\pm$48\\
31–-35&0.78&655$\pm$93&772$\pm$50\\
35–-41&0.76&711$\pm$95&708$\pm$50\\
41–-46&0.80&603$\pm$88&767$\pm$47\\
46–-52&0.74&570$\pm$110&755$\pm$56\\
52–-59&0.80&607$\pm$86&718$\pm$45\\
59–-67&{\bf 0.66}&{\bf 869$\pm$98}&{\bf 554$\pm$51}\\
67–-76&{\bf 0.54}&{\bf 1060$\pm$120}&{\bf 469$\pm$59}\\
76–-90&{\bf 0.43}&{\bf 960$\pm$190}&{\bf 452$\pm$89}\\
\hline
\label{tab:t-correlation}
\end{tabular}
\end{table}

\begin{figure}[ht]
\includegraphics[width=1.0\columnwidth,clip=]{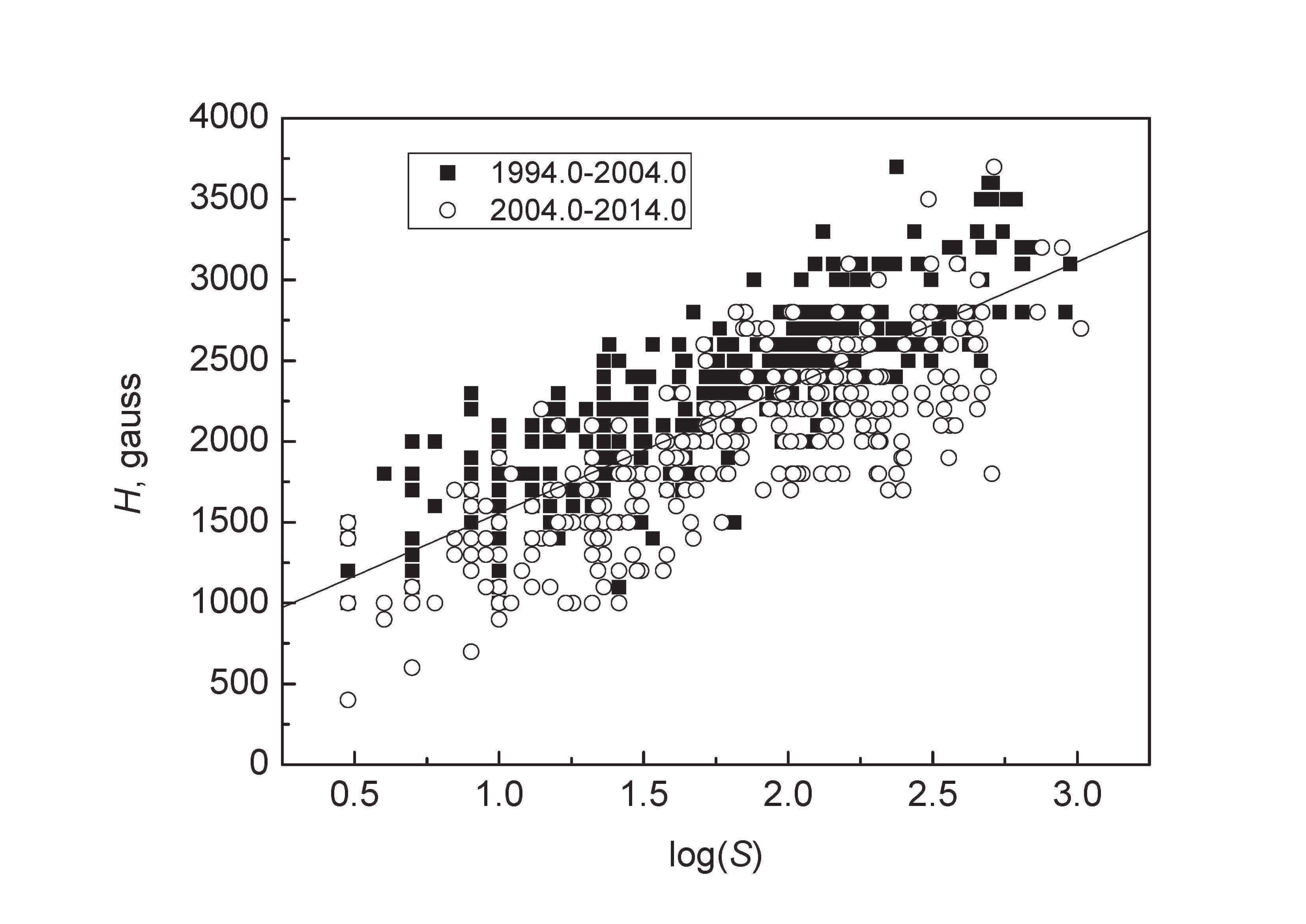}
\caption{Maximum field strength (Gauss) as function of area of sunspot (in units of millionth of solar hemisphere). Filled symbols correspond to 
1994.0--2004.0 observations, and opened symbols are for 2004.0--2014.0. Solid line shows least-square fit by 
Equation \ref{eq:H-log}.}
\label{fig:H-log}
\end{figure}

\section{Changes in Relation between Magnetic Field Strength and Sunspot Area}

Based on the results described in Section \ref{sec:func}, we limit our following investigation of
$H - \log S$ dependence to sunspots situated in central part of solar disk ($\theta \le$ 14$^\circ$,
$HS2$ data set, see Section \ref{sec:data}). Figure \ref{fig:H-log} shows scatter plot of $H$ vs. $\log S$ and its 
least-square fit $ H = (778\pm46) + (778\pm25) \times \log S$. r$_{\rm P}$=0.78. For the same sunspot areas, 
magnetic field strengths 
from early period (1994--2004, filled squares in Figure \ref{fig:H-log}) appears to be systematically higher as compared with 
the later period (2004--2014, open circles  in Figure \ref{fig:H-log}). There were no changes in observing procedure of measuring
sunspot magnetic fields during these two periods (Dr. Olga Gopasyuk, private communication, 2014). As an additional 
verification we compared 
measurements of sunspot field strengths in CrAO and MWO. For that we selected 100 measurements of the same sunspots
observed by both observatories on the same day (50 sunspots from 1994--2004 and 50 spots from 2004-2014). Mean difference
in measured field strengths $\Delta$ H$_{1994-2004}$ = 12$\pm$252 G and $\Delta$ H$_{2004-2014}$=76$\pm$375 G,
which further indicates that there were no significant  systematic changes in CrAO measurements over this period of time.
A comparative analysis of CrAO and MWO sunspot field strength measurements recently published by Lozitska et al. (2015)
also found a good agreement between the two datasets. 
In our CrAO--MWO comparison, we noted a weak tendency for CrAO measurements to exhibit lower field strengths in 
sunspots located 
closer to solar limb even thought the field strengths measured in the same spot when it was near disk center were
close to MWO measurements. We speculate that this tendency can be explained by higher level of scattered light 
in CrAO observations. Nevertheless, since $HS2$ dataset is limited to sunspots near disk center, the above mentioned
tendency has no effect on our conclusions. 

\begin{figure}[h!]
\includegraphics[width=1.0\columnwidth,clip=]{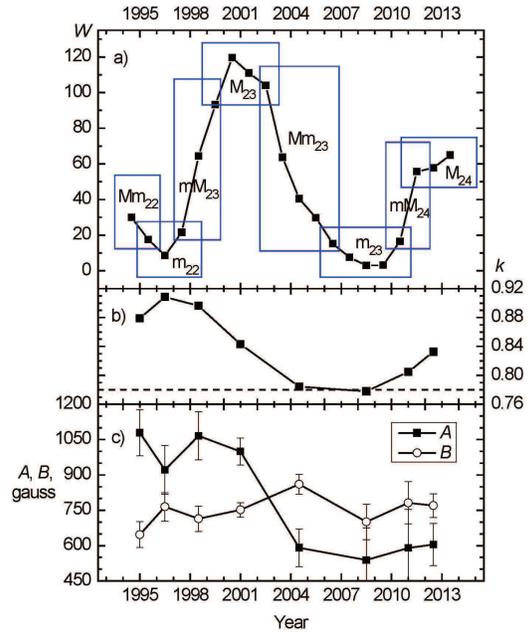}
\caption{International sunspot number (a) with boxes outlining subsets of data corresponding to different phases of solar cycle:  minimum (m), maximum (M), rising phase (mM) and declining phase (Mm). In all cases, subscripts refer to a numbered cycle ({\it e.g.},
m$_{23}$ corresponds to minimum between cycles 23 and 24, and mM$_{23}$ refers to rising phase of cycle 23. Two lower panels
show (b) changes in correlation coefficients between $H$ and $\log S$  and (c) variations
in  $A$ and $B$ coefficients (as in Equation \ref{eq:H-log}).}
\label{fig:m-M}
\end{figure}

Now, we can consider possible changes in $H - \log S$ dependence with phase of solar cycle. To do that, we divided $HS2$ data set 
into subsets representing different phases of sunspot cycle ({\it i.e.}, cycle minimum (m), maximum (M), rising phase (mM) and declining phase (Mm)). Graphical representation of each subset is given in Figure \ref{fig:m-M}a. The correlation coefficient 
between $H$ and $\log S$ varies 
with sunspot cycle from about 0.91 at minimum of cycle 22 to 0.78 at minimum of cycle 23. Changes in $B$ coefficient suggest that 
steepness of $H - \log S$ dependence does not change systematically between declining phase of cycle 22 and rising phase
of cycle 24. Changes in $A$ coefficient indicate a clear offset between Mm$_{22}$--M$_{23}$ and Mm$_{23}$ -- M$_{24}$
periods. This offset in $A$ coefficient is in agreement with the offset shown in Figure \ref{fig:H-log} between 
cycle 23 and cycle 24 
observations. Such changes maybe interpreted  as a systematic decrease in magnetic flux density in sunspots of the same area.

\begin{figure}[h!]
\includegraphics[width=1.0\columnwidth,clip=]{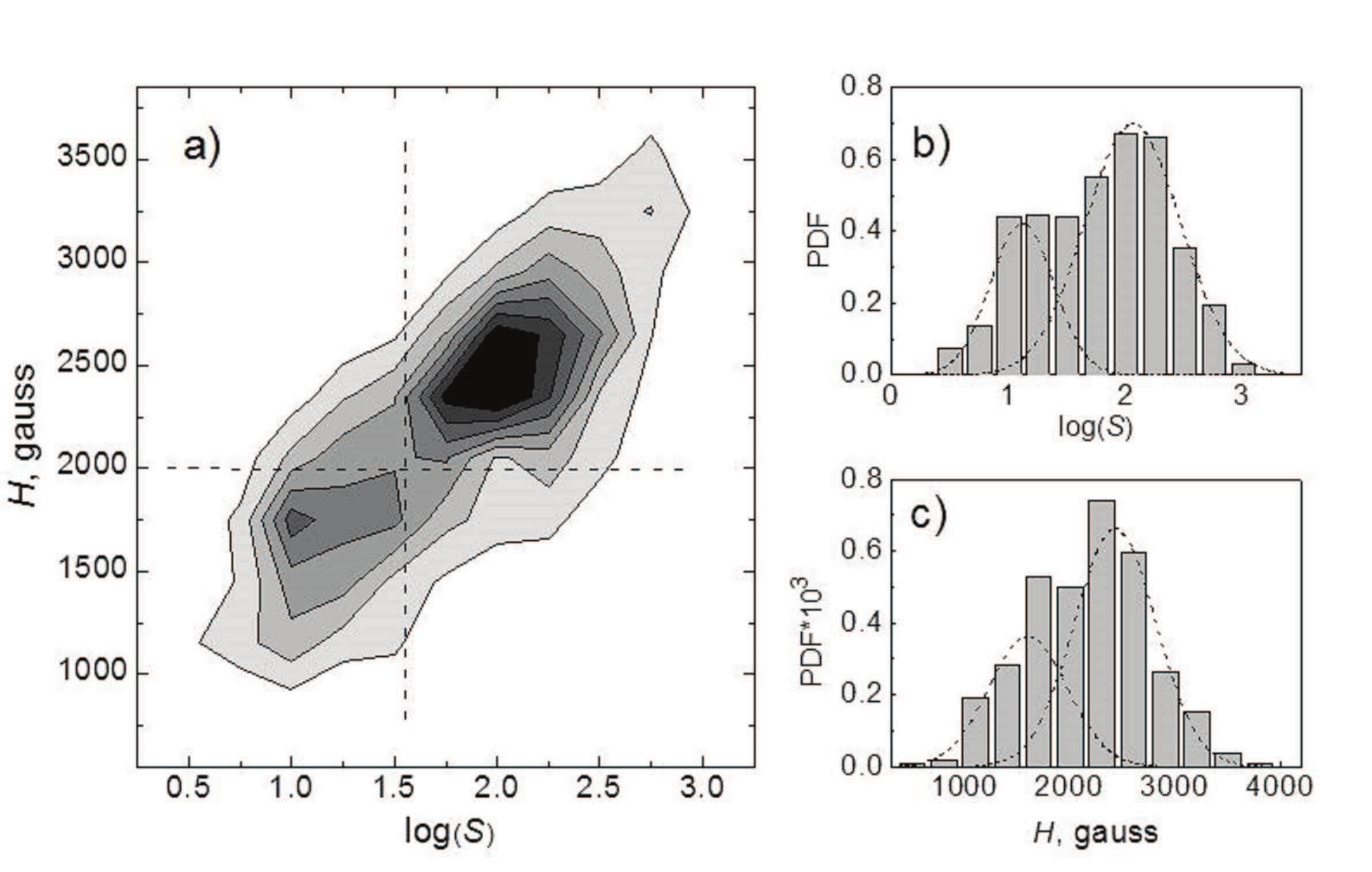}
\caption{Probability distribution functions (PDF) for sunspot area (b) and umbral field strength (c) for $HS2$ data set. Dashed lines
delineate two components contributing to each histogram. Panel (a) shows combined 2D PDF of two parameters. Vertical and 
horizontal dashed lines indicate approximate division between two components (small/weak and large/strong sunspots).}
\label{fig:2d}
\end{figure}

\begin{figure}[h!]
\includegraphics[width=.5\columnwidth,clip=]{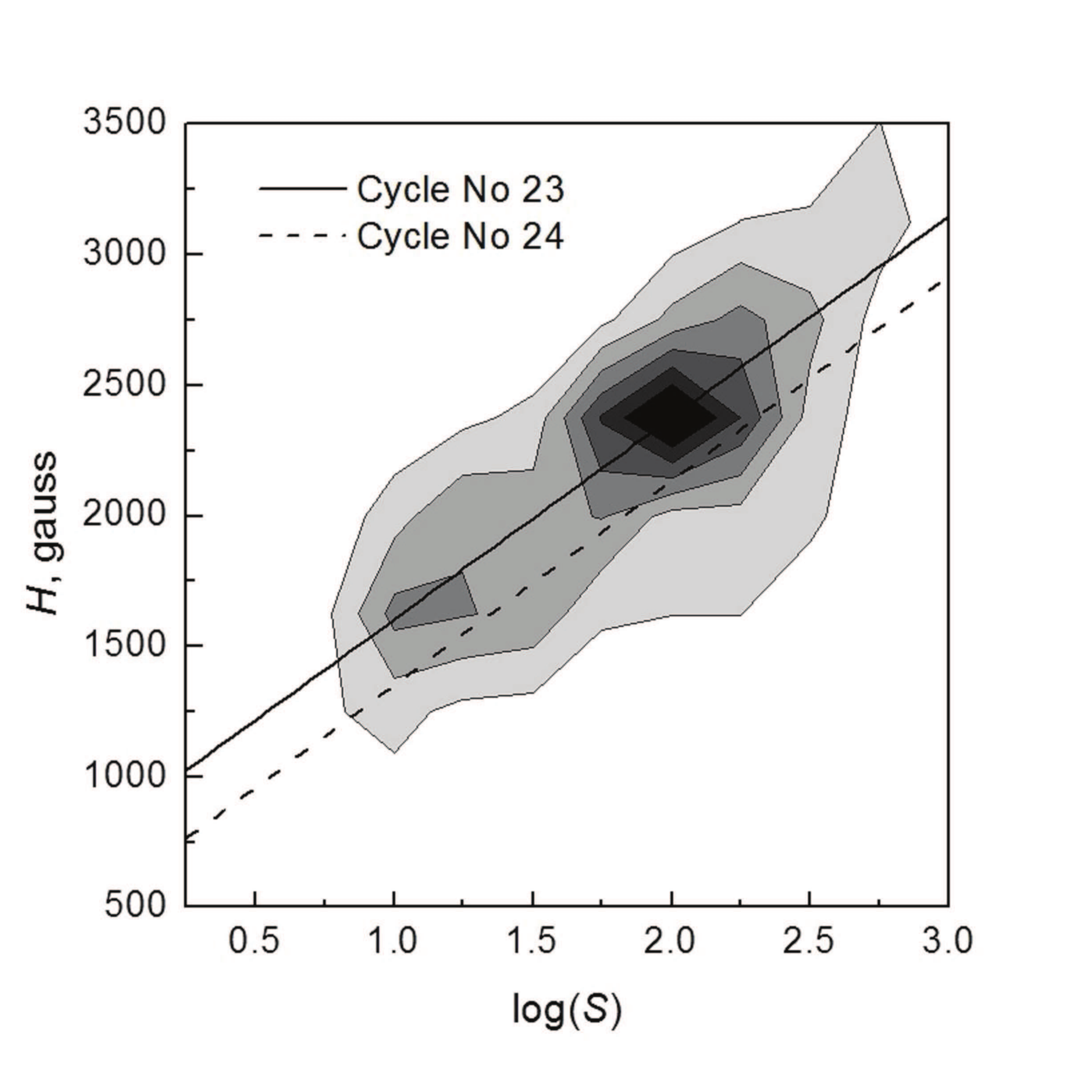}\includegraphics[width=.5\columnwidth,clip=]{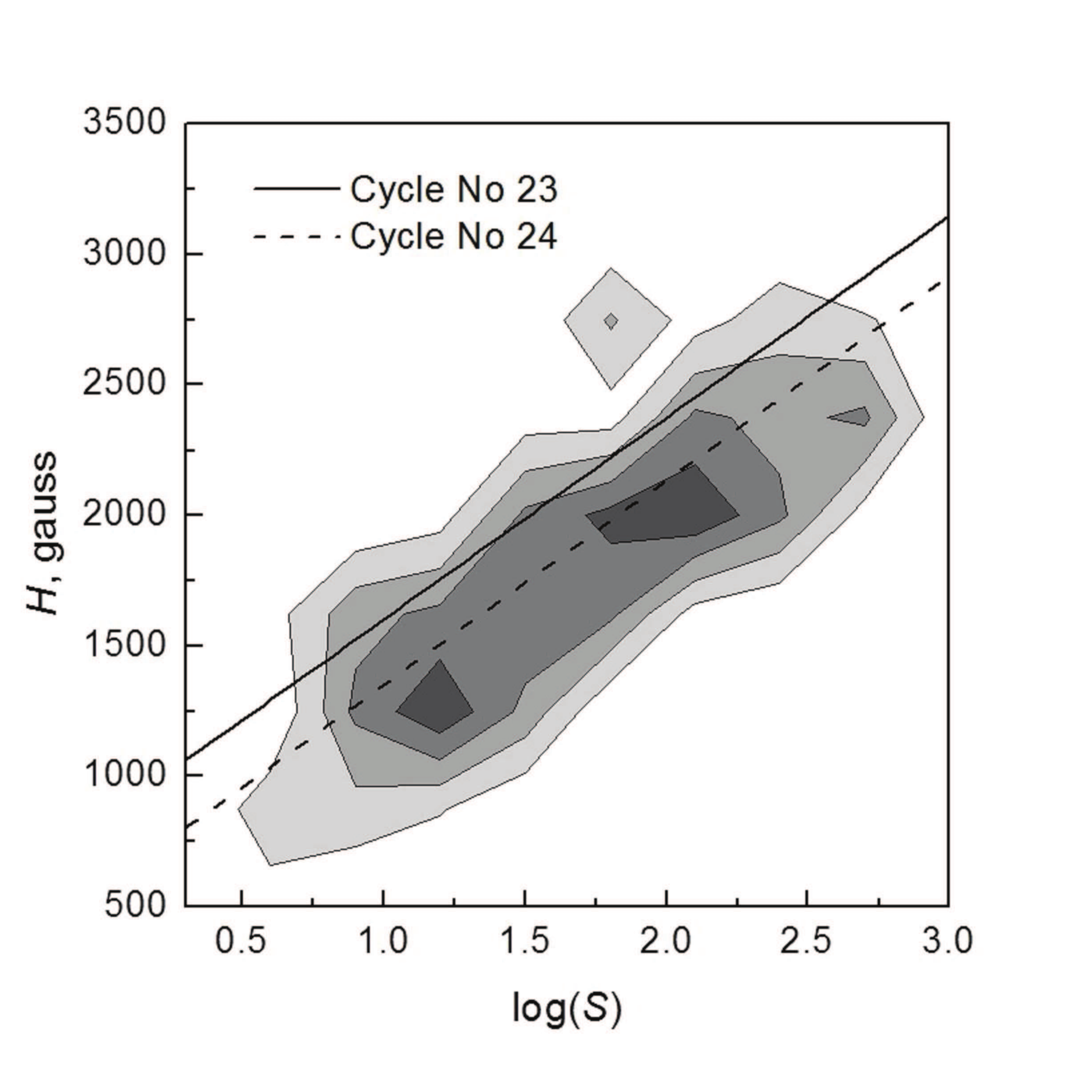}
\caption{2D probability distribution functions computed separately for cycle 23 (left) and cycle 24  (right). 
Solid line indicates 
first degree polynomial fit to $H$ vs. $\log S$ PDF for 
cycle 23, 
and dashed line shows similar fit to cycle 24  
data.}
\label{fig:2-2d}
\end{figure}

\begin{figure}[h!]
\includegraphics[width=1.0\columnwidth,clip=]{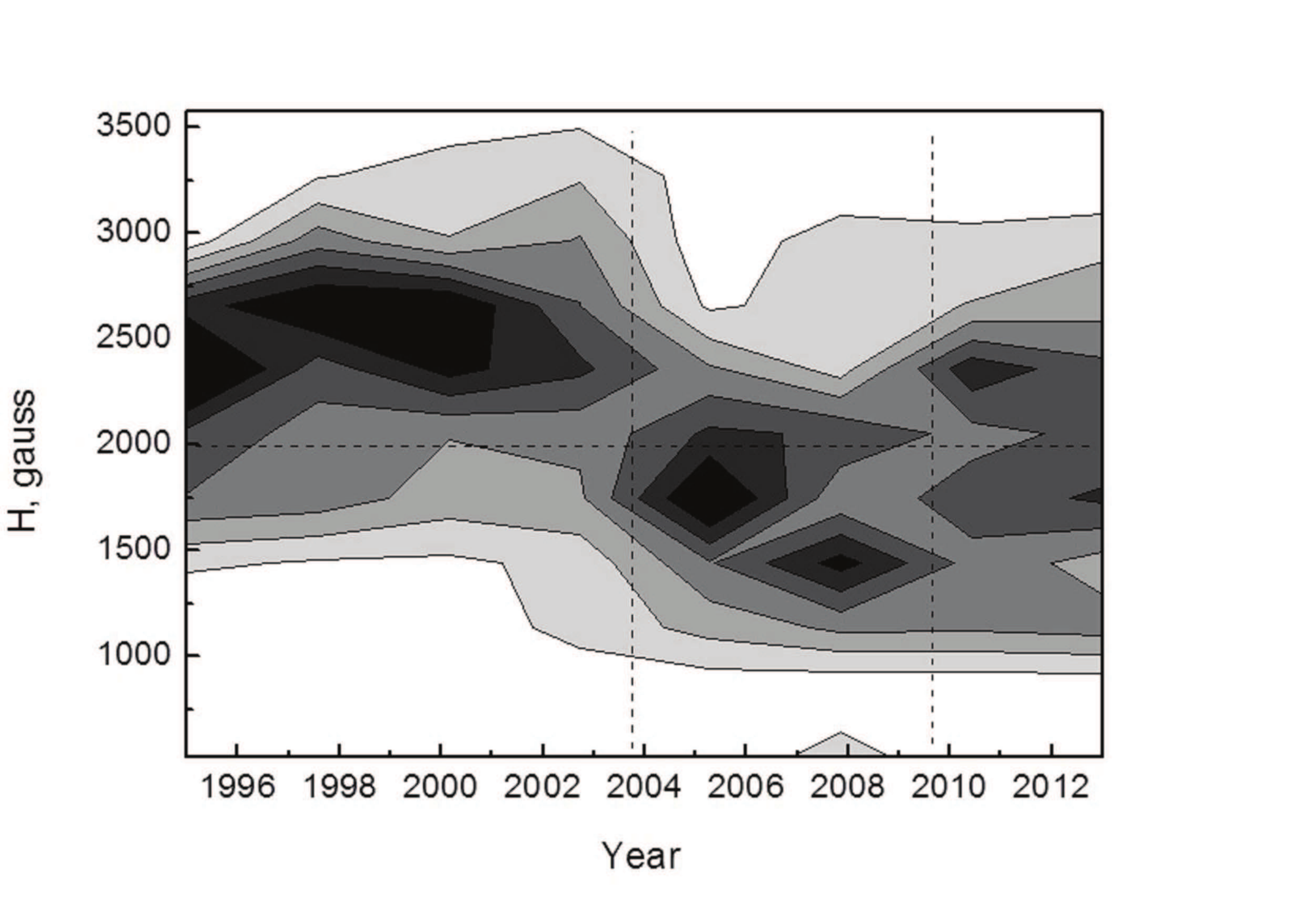}
\caption{A stack-plot showing PDFs for umbral magnetic field computed for each year in our $HS2$ data set. 
Vertical dashed lines mark the period of lowest 
correlation between magnetic field strength and sunspot area.}
\label{fig:long-term}
\end{figure}

\section{Bimodal distribution of sunspots}

Nagovitsyn et al. (2012) have shown that distribution of sunspot areas can be represented by a composite of 
two log-normal distributions, corresponding to "small'' and "large'' sunspots. Recent study by Mu{\~n}oz-Jaramillo et al. (2015) 
also show that the distribution of sunspot areas can be deconvolved on two components corresponding to sunspots of small and 
large areas. Since the sunspot magnetic field strength and their areas correlate with each other, one would expect to see bimodal distributions in both these parameters, which is, indeed, the case (see, Figure \ref{fig:2d}). In agreement with $H - \log S$ 
dependence, the sunspot areas are distributed log-normally and sunspot field strengths follow normal distribution. Both 
parameters show bimodal distribution (Figure \ref{fig:2d} right) with two distinct populations corresponding to small sunspots 
with weaker magnetic fields and larger sunspots with stronger fields. 

This change (in relative amplitudes of two contributing distributions) is clearly seen in Figure \ref{fig:2-2d}. In cycle 23, the distribution is skewed towards large sunspots with stronger
field (Figure \ref{fig:2-2d}, left). In cycle 24, large sunspot (stronger field strength) component of PDF
decreases considerably (Figure \ref{fig:2-2d}, right; 
compare contours in PDF for peaks corresponding to small and large sunspots around about $\log S = 1.2$ and $\log S = 2.1$),
and reaches about the same amplitude as small area (weaker field strength) component. 
The  change in relative contribution of sunspots with strong and weak field strengths is present in distributions built for annual intervals throughout the cycles 23 and 24 (Figure \ref{fig:long-term}). During 
most of cycle 23, sunspots with stronger field strength prevail, while near the minimum of cycle 23, a relative contribution of
sunspots with weaker fields increases. Cycle 24 shows a modest enhancement of contribution from sunspots with stronger fields, 
although the contribution of weaker sunspots is also clear (Figure \ref{fig:long-term}). 
  
This change in relative contribution of small and large sunspots is also present in a composite data set of sunspot 
group areas from Royal Greenwich Observatory (RGO) and US AF SOON network. For years around solar maxima, the distribution is skewed towards sunspots with larger areas with a peak of sunspot areas of about 400-500 $\times$ 10$^{-6}$ of solar hemisphere (SH). At solar minima, the mean of distribution is shifted towards smaller areas without a clearly defined peak.

\begin{figure*}
\includegraphics[width=2.0\columnwidth,clip=]{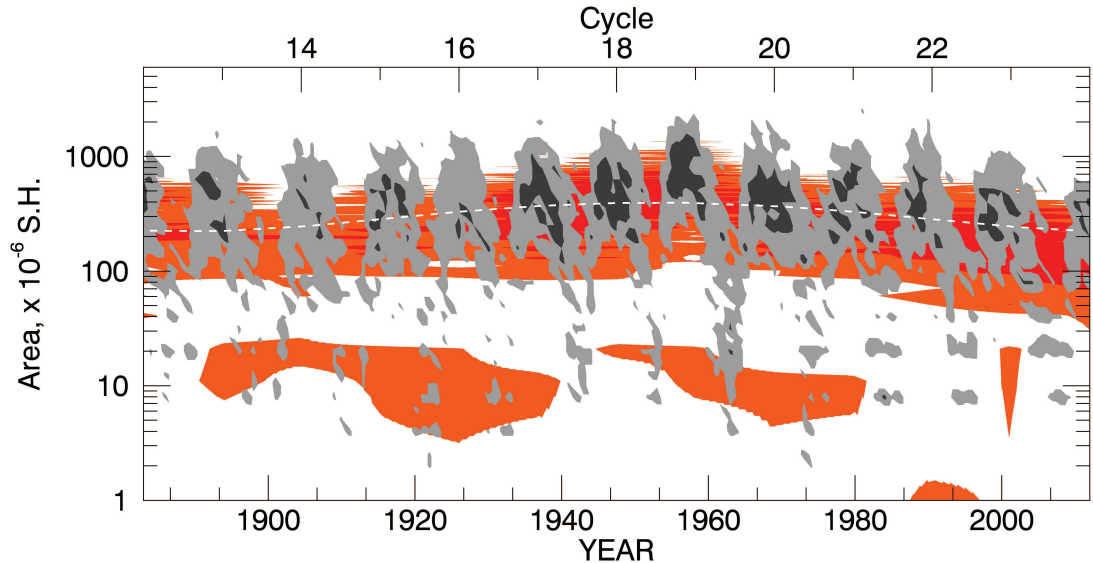}
\caption{A stack-plot of distributions of sunspots by their group area for the entire RGO--SOON data set.  The orange color corresponds to cycle averages, and gray scale shows the annual averages. A cross-section of this plot in the vertical direction represents individual distributions of sunspots for each year (cycle) normalized by total number of spots in this year (cycle). Shaded areas correspond to parts of distributions above a fixed threshold ($>$ 0.3 for light gray and $>$ 0.6 for dark gray). Long-term variations are more evident in solar cycle averages (orange). The white dashed line corresponds to a sine function fitted to the data.}
\label{fig:rgo}
\end{figure*}

Plotting the annual distribution of sunspot areas in RGO--SOON data reveals two major trends  (Figure \ref{fig:rgo}). First, the mean sunspot areas show a clear variation with the 11-year solar cycle (with sunspots having a tendency for larger areas around maxima of the 11-year cycle and smaller areas during solar minima). Second, the mean of (annual or cycle) distributions shows a tendency for a slight increase from the beginning of data set (1874) to mid-1950 followed by a gradual decrease from mid-1950 to end of 2013.

Nagovitsyn et al. (2012) interpreted the solar cycle variation of sunspot field strengths and the presence of two 
distinct components in sunspot area distributions in the framework of two dynamos, 
with smaller sunspots forming at more shallow depths, and larger sunspots forming deeper in the solar convection zone. Fitting the distribution of sunspot areas by a combination of Weibull and lognormal functions also supports the notion that 
small and large sunspots may be the result of different dynamo processes  (Mu{\~n}oz-Jaramillo et al. 2015). Thus, the 11-year variation in areas of sunspots can be explained by changes in relative contribution of deep seated and shallow dynamos.
A long-term trend is more apparent in the distribution of sunspot areas integrated over each solar cycle (Figure \ref{fig:rgo}, orange color). This tendency is further outlined by a sine function fitted to solar-cycle distribution of sunspot areas (dashed curve in Figure \ref{fig:rgo}). The shape and amplitude of this curve is in agreement with behavior of sunspot field strength proxy derived in Pevtsov et al. (2014). The fitted curve exhibits a broad minimum at about cycles 12--14 and a broad maximum in about cycles 18--19. It roughly coincides with 90-100 year (Gleissberg) cycle.  Next minimum of this fitted sine function is in the year 2015. We note that if the recent long-term changes in solar activity are due to the Gleissberg cycle, their minimum should occur later in 
the current cycle 24.
While this conclusion is based on the assumption that the long-term variations in Figure 6 are due to Gleissberg cycle, we note that 
similar conclusions were recently drawn by several other authors. For example, Zolotova and Ponyavin (2014) found that the activity in cycle 23 was similar to one in cycles prior to the Dalton and Gleissberg-Gnevyshev minima. Based on the level of the heliospheric 
magnetic field in solar minimum between cycles 23 and 24, Janardhan et al (2015) predicted that cycle 25 will be slightly weaker
as compared with cycle 24. Shepherd et al. (2014) predicted that the maximum sunspot number in cycle 25 to be only about 80\% of that in cycle 24.

\section{Discussion}

The results presented in this article support previous findings that sunspot areas and maximum field strengths are closely related.
Still, the exact functional dependency cannot be determined on the basis of statistical arguments alone; all three 
functional dependencies that we analyzed are similar to each other in a statistical sense. On the other hand, the strength of relation between sunspot areas and their field strength depends on location of sunspot on solar disk. Correlation between $S$ and $H$  decreases dramatically for heliocentric angles larger then 60$^\circ$, which could be explained by the increase in contribution of horizontal component of magnetic field and/or some instrumental effects such as (for example) increased contribution of scattered light affecting the magnetic field measurements. This change in correlation between the
magnetic field in sunspots and their areas as function of heliocentric distance needs to be taken into consideration.
For sunspots in the central part of solar disk, sunspot area $S$ and maximum magnetic field $H$ show strong correlation (correlation
coefficients 0.78 -- 0.91). 

To investigate the change in relation between sunspot area and its umbral magnetic field, we employ $H = A + B \times \log S$ functional dependence. 
We find that $B$ coefficient (slope) does not change significantly between cycle 23 and cycle 24, but $A$ coefficient (offset)
changes between 1994--2003 and 2004--2014 periods. The overall distributions of sunspot areas and field strengths
appear to be similar in two cycles, but the median of the distribution of field strengths for cycle 24 is slightly lower as compared with
cycle 23. This shift can be interpreted as if sunspots in cycle 24 have slightly smaller field strengths as sunspots of the same area in cycle 23. Alternatively, these changes can be 
explained by changes in fractional contribution of small and large sunspots to total distribution of sunspots in a particular year 
(or cycle). A significant change in offset (A coefficient in  $H = A + B \times \log S$) between 1994--2001 and 2004--2014 periods is puzzling. This offset, however, is independent on how we parse the dataset. As an additional test, we compared H--$\log S$ scatter plots for rising phases of cycle 23 and cycle 24, and we do see the same asymmetry, with branch corresponding to cycle 23 data located slightly above the branch corresponding to cycle 24 data similar to Figure 1. Thus, we do see the same behavior for three subsets with different division on time periods: 1994--2003 vs. 2004--2014 (Figure 1), cycle 23 vs. cycle 24 (Figure 4), and rising phase of cycle 23 vs. rising  phase of cycle 24 (no figure is shown).

\begin{figure}
\includegraphics[width=1.0\columnwidth,clip=]{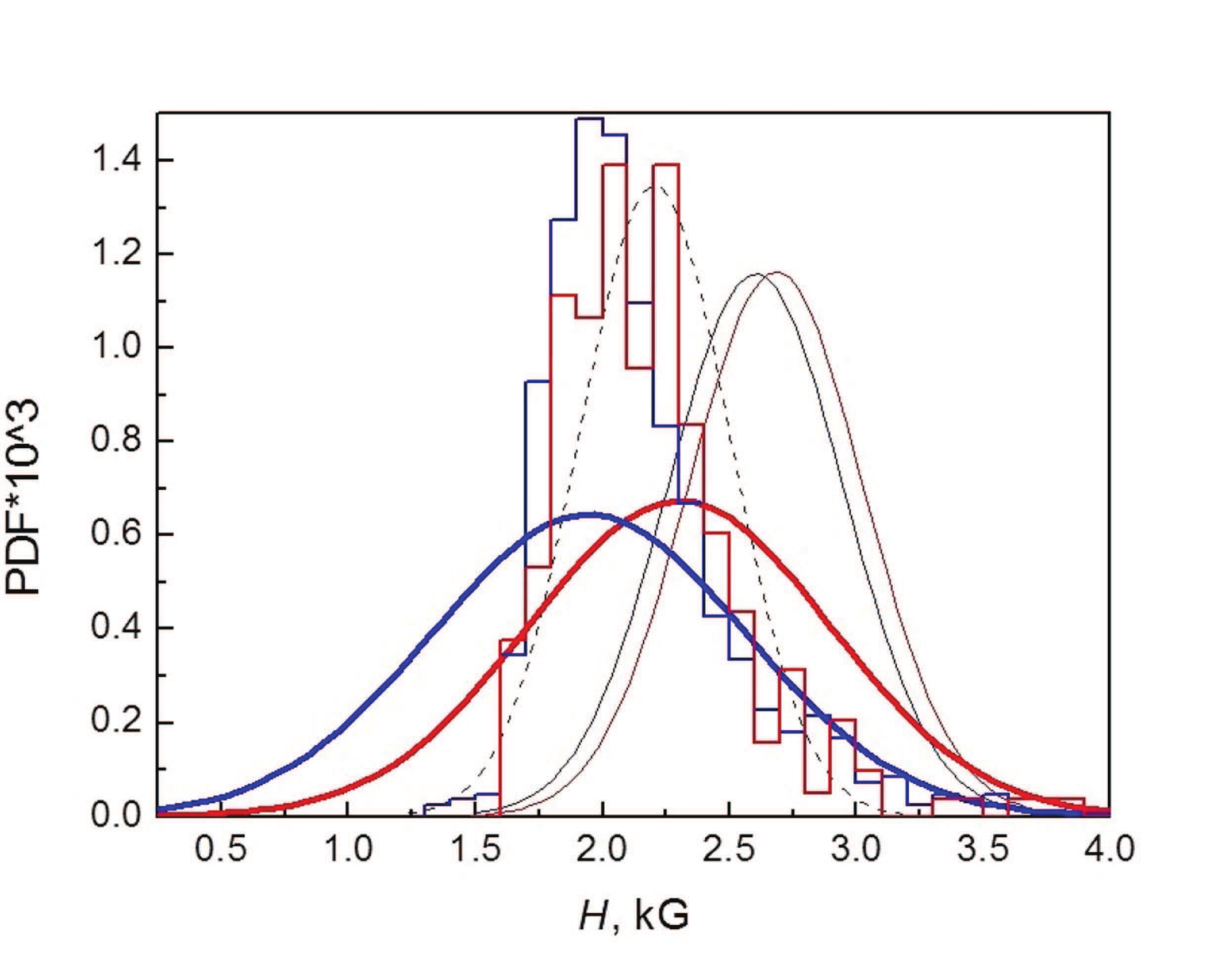}
\caption{Distributions of field strengths measured in sunspots in cycle 23 (red) and cycle 24 (blue). Thick lines show Gaussian fits to 
our data. Thin solid lines show Gaussian fits from Figure 5 in Rezai et al (2015), and two histograms are from Figure 9 in Schad (2014). Dotted line shows Gaussian fit to Livingston et al. (2012) observations from 2003--2007.}
\label{fig:gaussfit}
\end{figure}

Qualitatively, our finding that the magnetic fields in cycle 23 appear to be slightly stronger in comparison with cycle 24 seem 
to agree with the distribution of maximum field strengths shown in Figure 5 of Rezai et al. (2015). 
Figure \ref{fig:gaussfit} shows distributions of sunspot field strengths for cycles 23--24 reported by different authors. 
The mean of Gaussian distribution of maximum sunspot field strength as measured at CrAO in cycle 23 is about 2300 G, which is smaller than 2680 G found by Rezai et al. (2015) for the same cycle. Rezai et al. (2015) data are based on observations taken in infrared lines, which form deeper in the photosphere as compared with the Fe I 6302A (CrAO observations). Assuming that the gradient of the magnetic field in sunspots is about 1 G/km (e.g., Kotov 1970; Borrero \& Ichimoto 2011),  a 150--200 km difference in height of formation of spectral lines may account for 150—200 Gauss difference in the field strength. Even with that correction, the field strengths measured at CrAO are lower as compared with those reported by Rezai et al. (2015), which supports the notion that CrAO observations are affected by the scattered light. In addition, the Gaussian distributions of maximum field strengths in CrAO data are broader as compared with Rezai et al. (2015). This reflects the higher level of scatter (noise) in manual measurements. Similar to Rezai et al. (2015) the mean of Gaussian distribution of maximum sunspot field strength in cycle 24 is a few hundred Gauss smaller as compared with cycle 23, which 
is in an agreement with the results shown in Figures 1, 4, and \ref{fig:long-term}.

Schad (2014) used observations from Hinode taken in the photospheric spectral lines (Fe I 6301-6302A), and accordingly, the
mean of distributions reported by him are very close to our data (compare means of histograms and Gaussian fits shown by 
thick lines in 
Figure \ref{fig:gaussfit}). The width of distributions in Schad (2014) are much narrower as compared with our data, and more in line with Rezai et al (2015). Distribution of sunspot field strengths from Livingston and Penn measurements (cycle 23 only) show a narrow
width in agreement with Schad (2014) and Rezai et al (2015), but the mean of distribution is significantly smaller than in Rezai et al (2015), although the measurements were also taken in near IR forming in deep photosphere. Contrary to Rezai et al (2015) 
and our data presented here,  Schad (2014) did not find significant difference in mean field strengths between cycles 23 and 24.
However, this could be explained by the limited period this data set covers (November 2006 -- November 2012).
According to Figure \ref{fig:long-term}, the PDF of umbral fields shows the presence of stronger fields before year 2004; in 2006-2012 period (Hinode data used by Schad, 2014), the PDF exhibits the presence of a weaker component of field strength only.

By itself, the distribution of sunspot areas and their magnetic fields are bimodal with one component representing the
contribution of small sunspots with weaker fields and the other component representing large sunspots with stronger fields. The sunspot areas
follow the log-normal distribution, and the sunspot field strengths are distributed normally. Indirectly, this supports the  
 $H = A + B \times \log S$ as the most appropriate representation of functional dependence between $S$ and $H$.
Bimodal distribution of sunspots (by their area and field strength) may be interpreted as an indication of a dynamo, in which 
the generation of sunspots  of different size (and field strength) is spatially separated. Observationally, 
one component of bimodal distribution may be related to transitional sunspots and large pores, while the other component 
is related to mature sunspots. Distribution of sunspot umbral areas recently published by Cho et al. (2015) seems to 
support this idea.

Finally, sunspot area data exhibit the presence of long-term variations, which may be in phase with 90--100 year cyclic variations
of solar activity (Gleissberg cycle). The data suggest that the next minimum of this centennial cycle will occur later in cycle 24. 

\acknowledgements
Y.A.N. and A.A.O. acknowledge support by
the Russian Foundation for Basic Research 
(RFBR grant No 16-02-00090)
 and research program 
No. 7
of the Presidium
of Russian Academy of Sciences.
National Solar Observatory is operated by the Association of
Universities for Research in Astronomy (AURA), Inc. under a
cooperative agreement with the National Science Foundation.
This work utilizes synoptic data from Crimean Astrophysical Observatory (CrAO) and Kislovodsk Mountain 
Astronomical Station (KMAS). The authors thank CrAO and KMAS teams for their open data policy. We also 
thank Ms. J. Diehl for a careful proofreading of the manuscript. 
We acknowledge comments by the anonymous reviewer, which helped to improve the article.


\end{document}